\documentclass[12pt]{article}
\usepackage{amssymb}
\usepackage{amsfonts}
\usepackage{psfrag}
\usepackage{epsfig}
\usepackage{bbm}

\newcommand{\bcen}{\begin{center}}
\newcommand{\ecen}{\end{center}}
\newcommand{\brig}{\begin{flushright}}
\newcommand{\erig}{\end{flushright}}

\newcommand{\barray}{\begin{array}}
\newcommand{\eeqarray}{\end{eqnarray*}}
\newcommand{\beqarray}{\begin{eqnarray*}}
\newcommand{\earray}{\end{array}}
\newcommand{\bite}{\begin{itemize}}
\newcommand{\eite}{\end{itemize}}
\newcommand{\bmath}{\begin{displaymath}}
\newcommand{\emath}{\end{displaymath}}
\newcommand{\beq}{\begin{equation}}
\newcommand{\eeq}{\end{equation}}
\newcommand{\bea}{\begin{eqnarray}}
\newcommand{\eea}{\end{eqnarray}}
\newcommand{\bdm}{\begin{displaymath}}
\newcommand{\edm}{\end{displaymath}}

\newcommand{\as}{\alpha_s}

\begin{document}

\pagestyle{empty}

\vspace*{13mm}
 
\boldmath
\begin{center}
{\LARGE{\bf Hadronic mass and $q^2$ moments in $B \to X_u l \nu$
}}
\vspace*{4mm}
\end{center}
\unboldmath

\smallskip
\begin{center}
{\large{Giovanni Ossola}}  \\
\vspace{2mm}
{\sl Dipartimento di Fisica Teorica, Universit\`a di Torino,\\   
and INFN, Sez.~Torino, I-10125 Torino, Italy}\\
\vspace{1mm}
{\tt ossola@to.infn.it}
\vspace*{24mm}

\abstract{
We present OPE predictions for the hadronic mass
and $q^2$ moments in inclusive semileptonic charmless
$B$ decays, with a lower cut on the charged lepton 
energy and an upper cut on the hadronic invariant mass.
We include non-perturbative corrections through ${\cal O}(1/m_b^3)$ 
and perturbative contributions through ${\cal O}(\alpha_s^2 \beta_0)$.
We also investigate the range of the cut on the hadronic mass
for which the local OPE can be considered valid and give estimates 
of the residual theoretical uncertainty.
 
The study of these moments is important to constrain 
the effect of the Weak Annihilation (WA) contributions 
and the shape of the distribution function, 
providing a more precise inclusive determination of $|V_{ub}|$.
}
\vspace*{18mm}
\end{center}

\bcen
\emph{To appear in the proceedings of the \\
International Europhysics Conference on High Energy Physics\\
                 July 21st - 27th 2005, 
                 Lisboa, Portugal}
\end{center}

\section{Introduction and General Motivations}

The determination of the $V_{ub}$ element of the CKM matrix, together with
a sensible estimate of the theoretical uncertainties involved in 
its calculation, is an important but challenging task. 
The magnitude of $V_{ub}$, combined with
$V_{cb}$, determines the length of the left side of the Unitary Triangle, 
allowing for a stringent test of the Standard Model and 
a providing a powerful tool to detect hints of new physics \cite{utfit1}.
While $V_{cb}$, thanks to recent developments, has reached a theoretical 
uncertainty below 2\%, the situation of $V_{ub}$ is much worse, 
with uncertainties about 10\% \cite{hfag,revs}. Therefore improvements are needed here.

The difficulties related with the determination of the $V_{ub}$ from inclusive
semileptonic decays are well known. 
Inclusive decay rates can be calculated using a double expansion in 
$\alpha_s$ (parton model) and $\Lambda_{QCD}/{m_b}$ (Heavy Quark Expansion):
\beq
\Gamma(\bar B\to X_u \ell \bar\nu) = \frac{G_F |V_{ub}|^2 m_b^5}{192 \pi^3}
\left[ 1+ \sum_{i,\, n} C_{(i,n)}\, \left(\frac{\alpha_s}{\pi}\right)^{i} 
\left(\frac{\Lambda_{QCD}}{m_b} \right)^n \right]\,\, .
\eeq
The measurement of  $\bar B \to X_u\ \ell \bar\nu$ requires experimental 
cuts to
suppress the large $\bar B \to X_c\ \ell \bar\nu$ background. 
Unfortunately, the 
introduction of kinematic cuts restricts the phase space available for the
$\bar B \to X_u\ \ell \bar\nu$ decay, spoiling the global 
properties of the OPE.

Several possibilities have been explored in the literature \cite{cuts1}, 
involving cuts in the hadronic invariant mass ($M_X$), lepton energy ($E_l$), 
leptonic invariant mass ($q^2$), light-cone component of the hadronic 
four-momentum ($P_+$), and combinations of them. Each of these cases 
requires different techniques and
a careful estimates of the uncertainties involved.

In this talk, I will present a calculation of the moments in 
the hadronic invariant mass of the $\bar B\to X_u \ell \bar\nu$ 
distribution \cite{gou}, and try to suggest the 
reason for its usefulness.
Of course, as soon as the measurements of such moments will be available, 
one should verify their consistency with moments of 
$B\to X_c \ell \bar\nu$ and $B\to X_s \gamma$ in the OPE framework.
The present combined fit \cite{fits} of all moments of inclusive B decays 
(hadronic mass, lepton energy) shows very good consistency of 
the available data and allows for the determination of the $b$ quark mass
and expectation values of the dominant power suppressed operators.      

However the main motivations for our analysis comes from recent 
experimental developments. The high statistics accumulated at 
the B-factories by BaBar and Belle allows for
the measurement, based on fully reconstructed events, of hadronic 
invariant mass distributions in inclusive $\bar B\to X_u \ell \bar\nu$. 
This new generation of analysis can discriminate between 
charmless events and charmed background, even for 
relatively high values of hadronic invariant mass $M_X$.
For instance, BaBar has been able to measure
invariant mass distribution and first two moments with promising 
accuracy \cite{babar}. 
However, the experimental error is much smaller if one applies a 
cut on the hadronic mass $M_X^{\rm cut}$ close to the kinematic boundary 
for charm production.
On the other hand, theoretical calculations of inclusive 
B decays rely on the OPE, whose convergence is spoiled by severe cuts.
A possible solution could be to raise $M_X^{\rm cut}$ just enough to suppress 
non-perturbative effects that cannot be accounted for by the 
OPE ($B$ meson distribution function effects).
I will show a simple example in the Discussion (Sect.~4), 
but before let me briefly describe the content of our calculation (Sect.~2) 
and succinctly discuss the theoretical uncertainties involved (Sect.~3).

\section{Hadronic mass moments}

We determine the normalized integer moments of the squared invariant mass 
\beq
\langle M_X^{2n}\rangle = \frac{\int d  M_X^2 \ M_X^{2n}\  d\Gamma/d M_X^2}
{\int d  M_X^2\  d\Gamma/d M_X^2}
\eeq 
and in particular the central moments defined as
$$ U_1= \langle M_X^{2}\rangle \,\, ,\,\, U_{2,3}=\langle
\left( M_X^{2}-\langle M_X^{2}\rangle\right)^{2,3}\rangle\,\, . $$
The physical hadronic invariant mass is related to parton level quantities by
$$M_X^2=\bar \Lambda^2 +2m_b \bar \Lambda E_0 + m_b^2  s_0 \,\, ,$$ 
therefore 
we can express the moments of $M_X^2$ in term of combinations of moments 
of parton energy $E_0$ and invariant mass of the parton $s_0$.
Using the same approach, we can easily obtain also the moments 
of the distribution in $q^2$.

The building blocks that allow to construct both $M_X^2$ and $q^2$ moments 
are:
\begin{eqnarray} \label{bb}
{\cal M}_{(i,j)} &=& \frac1{\Gamma_0}
\int  d E_0\, d s_0 \,d E_\ell 
\ {s_0}^{i} \ 
 {E_0}^{j} \ \frac{d^3 \Gamma}{d E_0 \,d s_0 \,d E_\ell}  \\ \nonumber
&=& T_{(i,j)} +  \frac{\mu_\pi^2}{m_b^2} B_{(i,j)}
 + \frac{\mu_G^2}{m_b^2} C_{(i,j)} 
+ \frac{\rho_D^3}{m_b^3} D_{(i,j)} + \frac{\rho_{LS}^3}{m_b^3} E_{(i,j)} 
+ \frac{\alpha_s}{\pi} A_{(i,j)}^{(1)}
+\frac{\alpha^2_s\ \beta_0}{\pi^2} A_{(i,j)}^{(2)}\, . 
\end{eqnarray}
This expression does not include non-perturbative terms of 
${\cal O}(1/m_b^4)$, ${\cal O}(\as / m_b^2)$, and perturbative corrections 
of ${\cal O}(\alpha_s^2)$. 

The moments of $E_0$ and $s_0$ and of their product are obtained 
in the local OPE and are expressed in terms of the heavy quark parameters.

We employ  the ``kinetic'' scheme \cite{kin} (Wilsonian scheme with a hard factorization 
scale $\mu \simeq 1$ GeV). 
We start from on-shell expressions, and express
the on-shell parameters in terms of the  $\mu$-dependent ``kinetic'' 
parameters: kinetic mass of the $b$ quark $m_b(\mu)$,  
kinetic expectation value $\mu_\pi^2(\mu)$, and 
Darwin expectation value $\rho_D^3(\mu)$. 
At $\mu\!\to\!0$ one recovers the results of the on-shell scheme.

In the case of charmless decay, we introduce an additional parameter 
$X_\mu \equiv 8 \ln m_b^2/\mu_{4q}^2$.  
It is well known \cite{gremm, blok} that the coefficient 
function $D_{(0,0)}$ (see Eq.~\ref{bb}) 
has a logarithmic divergence as $m_q\!\to\!0$, 
i.e. contains a term proportional to $\ln\ ( m_b^2/m_q^2)$.
This problem is solved when we include the contribution of the 
Weak Annihilation (WA) operator. In fact, 
a one-loop penguin diagram that mixes the four-quark operator into  
$\rho_D^3$  replaces 
$\ln{m_u^2}/{m_b^2}$  by $\ln{\mu_{4q}^2}/{m_b^2}$, where
$\mu_{4q}$ is the normalization point of the WA operator.
Varying the parameter $X_\mu$ allows us to estimate the effects of WA.
The associated variation in the results for the moments accounts 
for the non-valence piece in the expectation value of the WA operator 
(flavor singlet WA). This contribution does not distinguish between 
$B^+$ and $B^0$ (non-singlet WA effect).

The calculation is implemented in a FORTRAN 
code\footnote{Available upon request from the authors of Ref.~\cite{gou}.}
and includes non-perturbative corrections ${\cal O}(1/m_b^2)$ \cite{1mb2} 
and  ${\cal O}(1/m_b^3)$ \cite{gremm},
and perturbative corrections ${\cal O}(\alpha_s)$ and 
${\cal O}(\alpha_s^2\ \beta_0)$ \cite{pert}.
Partial checks have been performed to reproduce previous results \cite{dfn}.

\section{Theoretical Uncertainties}

There are several sources of theoretical uncertainty in our predictions 
for the moments. Let me briefly list them and describe the methods used to 
estimate their size. We should consider:
(i) Uncalculated  ${\cal O}(\alpha_s^2)$ and 
 perturbative corrections to the Wilson coefficients; 
(ii) Missing ${\cal O}(1/m_b^4)$ non-perturbative effects;
(iii) Error from the scale in $X_\mu$;
(iv) WA contributions;
(v) Effects of Fermi Motion.

We estimate the effect of missing higher order corrections (i) and (ii), 
by varying the parameters $\mu_\pi^2$, $\mu_G^2$, $\alpha_s$, $\rho_D^3$, 
$\rho_{LS}^3$, and $m_b$ in an uncorrelated way. Alternatively, we can 
estimate the size of missing higher orders comparing the results 
obtained for the moments using different rearrangements of 
non-perturbative corrections. 
As I mentioned previously, the effects of (iii) and (iv) 
are closely related \cite{gou}: varying $\mu_{4q}$ over a 
reasonable interval, we estimate the effect of the flavor singlet 
WA. Those contributions are concentrated at maximal $q^2$
(small hadronic invariant mass) and are therefore 
suppressed in the moments of $M_X^2$.
Finally, we estimate the effect of Fermi motion (v), 
introduced by the $M_X$ cut, 
by smearing the tree-level differential rate with an 
exponential distribution function. 
This estimate depends on the functional form adopted, for instance 
an  exponential form leads to more conservative estimates
than a Gaussian ansatz.
While find relatively small effects for the first moment, the second and 
the third one are more sensitive. A more complete and satisfactory 
treatment of these effects is required, and it is 
currently under investigation.

\section{Discussion and Conclusions}

As an example of the possibilities offered by the analysis based on the 
moments, we present an estimate of the error on a the determination of $m_b$ 
based on $U_1$ only. We consider a mild  $M_X^2 <$5.6 ${\rm GeV}^2$ cut, 
that has the 
advantage of removing the experimentally most poorly known region 
of phase-space for charmless decays, reducing the experimental error by 
almost a factor two, respect to the case without the cut \cite{tack}.
The experimental uncertainty, considering the statistics that can be 
accumulated in the near future, can be estimated as 
$\delta U_1^{\rm exp} = 11\% $ \cite{tack}.  
On the theoretical side, considering the sources of uncertainty listed above, we obtain $\delta U_1^{\rm th} = 9\%$ at $M_X^{\rm cut}=$2.5~GeV\cite{gou}.
The combined effect of theoretical and experimental uncertainties, 
together with the use of the linearized formulas of Ref.~\cite{gou}, roughly
leads to $\delta m_b \simeq$ 80~MeV.

To summarize, we presented predictions for the first three moments 
of the hadronic invariant mass distribution in  $\bar B\to X_u \ell \bar\nu$,
that includes corrections through $O(\alpha_s^2\beta_0)$ and $O(1/m_b^3)$ 
and cuts on the lepton energy and on the invariant hadronic mass.
The calculation is implemented in a FORTRAN code, available upon request.
The work is in progress: we are currently improving our code, allowing 
all possible combined cuts (including $q^2$ and light-cone variables). 
A lot is still to be done, in order to constrain WA (singlet/non-singlet) and
better understand Fermi motion effects.
An analysis based on moments with high $M_X^{\rm cut}$ is challenging for 
experiments, but reduces the theoretical uncertainties 
related with OPE predictions.  The preliminary results are promising.

\vspace{0.6cm}

\noindent {\bf Acknowledgments} \\
Work done in collaboration with 
Paolo Gambino and Nikolai Uraltsev, supported in part by the EU grant
MERG-CT-2004-511156 and by MIUR under contract 2004021808-009.

\end{document}